%% file: main.tex
\documentclass[10pt,conference]{IEEEtran}
\IEEEoverridecommandlockouts
\usepackage{cite}
\usepackage{amsmath,amssymb,amsfonts}
\usepackage{algorithm}
\usepackage{graphicx}
\usepackage{textcomp}
\usepackage{subcaption}
\usepackage{xcolor}
\usepackage{booktabs}
\usepackage{svg}
\usepackage{siunitx}
\usepackage[noend]{algpseudocode}
\usepackage{braket}
\usepackage[hidelinks]{hyperref}
\usepackage{svg}
\usepackage{tikz}

\newcommand{\circled}[1]{%
\tikz[baseline=(char.base)]{
\node[shape=circle,draw,inner sep=2pt] (char) {#1};
}%
}

\def\BibTeX{{\rm B\kern-.05em{\sc i\kern-.025em b}\kern-.08em 
    T\kern-.1667em\lower.7ex\hbox{E}\kern-.125emX}}

\begin{document}

\title{Centralizing Task-based Approach to Quantum Network Control\\
\thanks{This material is based upon work supported by the U.S. Department of Energy, Office of Science, Advanced Scientific Computing Research (ASCR) program under contract number DE-AC02-06CH11357 as part of the InterQnet quantum networking project. R.H., J.C., and R.K. acknowledge support from the Argonne National Laboratory Directed Research and Development (LDRD) program.}
}

\author{
    \IEEEauthorblockN{Alexander Pirker\IEEEauthorrefmark{1},
    Robert J. Hayek\IEEEauthorrefmark{2},
    Alexander Kolar\IEEEauthorrefmark{2}\IEEEauthorrefmark{3},
    Igor Kadota\IEEEauthorrefmark{4}, 
    Joaquin Chung\IEEEauthorrefmark{2}, 
    Rajkumar Kettimuthu\IEEEauthorrefmark{2}
    }
    \IEEEauthorblockA{
        \IEEEauthorrefmark{1}
        Quantum Network Design GmbH, Innsbruck, Austria \\
        Email: alexander.pirker@qnetworkdesign.com \\
    }
    \IEEEauthorblockA{
        \IEEEauthorrefmark{2}
        Data Science and Learning, Argonne National Laboratory, Lemont, USA \\
        Email: \{rhayek, akolar, chungmiranda, kettimut\}@anl.gov \\
    }

    \IEEEauthorblockA{
        \IEEEauthorrefmark{3}
        Pritzker School of Molecular Engineering, University of Chicago, Chicago, USA \\
        Email: atkolar@uchicago.edu
    }
    \IEEEauthorblockA{
        \IEEEauthorrefmark{4}
        Department of Electrical and Computer Engineering, Northwestern University, Evanston, USA \\
        Email: kadota@northwestern.edu \\
    }
}

\maketitle

\begin{abstract}
For the last decade, layered stacks have dominated the way of reasoning about architectures for quantum networks. However, layered architectures impose stringent design and timing constraints on quantum networks, adding additional latency to the time required to serve an entanglement generation request. Moreover, increasing delays from the layered approach to network control causes degradation of state, effectively minimizing achievable fidelities. In this work we simulate a resource-centric, task-based approach to quantum network control by utilizing a centralized controller. Using the SeQUeNCe quantum network simulator, we implement the centralized controller which tracks quantum memory availability across all nodes, and schedules objectives in an offline fashion using a priority-based scheduler. We evaluate the performance of this controller on multiple topologies (bottleneck, grid, star, caveman) of significant scale, with varying reservation patterns; thereby we demonstrate the viability of the resource-centric task-based quantum network control framework for scaling. Our simulation results show that the caveman and grid topologies have a higher fraction of delivered requests with low delay compared to the star topology, but with a higher fraction of highly delayed requests as well. Furthermore, we find a linear shift of the CDFs in terms of queue size for all topologies depending on the reservation delay. More interestingly, we conclude that the CDFs of priority queues for the star topology converge fast into saturation for increasing request arrival rates, demonstrating together with the other results that the framework is robust for high load scenarios in quantum networks. 
\end{abstract}


\input{sections/introduction}
\input{sections/background}
\input{sections/architecture}
\input{sections/system_model}
\input{sections/methodology}

\input{sections/results}
\input{sections/conclusion}

\section*{Acknowledgment}
The authors would also like to thank Caitao Zhan for insightful discussion that lead to the implementation of the centralized controller in SeQUeNCe.

\bibliographystyle{ieeetr}
\bibliography{bibliography}

\section*{Government License}
The submitted manuscript has been created by UChicago Argonne, LLC, Operator of Argonne National Laboratory (``Argonne''). Argonne, a U.S. Department of Energy Office of Science laboratory, is operated under Contract No. DE-AC02-06CH11357. The U.S. Government retains for itself, and others acting on its behalf, a paid-up nonexclusive, irrevocable worldwide license in said article to reproduce, prepare derivative works, distribute copies to the public, and perform publicly and display publicly, by or on behalf of the Government. The Department of Energy will provide public access to these results of federally sponsored research in accordance with the DOE Public Access Plan. \href{http://energy.gov/downloads/doe-public-access-plan}{http://energy.gov/downloads/doe-public-access-plan.}

\end{document}

%% file: sections/introduction.tex
\section{Introduction} \label{sec:intro}
Quantum networks~\cite{van2014quantum} have become a popular subject of study due to their potential to enhance classical networks (e.g., quantum key distribution~\cite{ekert1991quantum} and high-precision clock synchronization~\cite{komar2014quantum}) as well as create new applications (e.g., blind quantum computation~\cite{broadbent2009universal} and distributed quantum sensing~\cite{zhang2021distributed}).
Recently, they have gained even more popularity as scaling to millions of physical qubits on a single chip remains a fundamental barrier to fault-tolerant quantum computing (FTQC)~\cite{quantum_interconnect,mohseni2026buildquantumsupercomputerscaling,eisert2025mindgapsfraughtroad}.
Thus, linking quantum processing units (QPUs) via quantum networks has been identified as a suitable path for scalability.
A longer term goal is to eventually create a quantum internet~\cite{kimble2008quantum,wehner2018quantum,kozlowski_rfc_2023} connecting quantum resources across the globe.

The main goal of quantum networks is to distribute quantum entanglement among remote parties, either bipartite entanglement (in terms of Bell-pairs) or multipartite entanglement (in terms of graph states).
To reason about a future quantum internet, researchers have proposed multiple architectures and protocol stacks.
In general, three different kinds of quantum network stacks have been proposed and pursued in the last decade~\cite{VanMeterStack1,VanMeterQRNA,WehnerStack1,WehnerStack2,WehnerStack3,Pirker_2019}. 
They all have in common a hierarchical organization into layers to segregate responsibilities in a quantum network, mirroring the success of the layered stack of the classical Internet.
However, these stacks substantially differ in the way that they organize the network functionality into layers.

Layered architectures impose stringent requirements on hardware devices, which include asynchronous operations enabled by quantum memories with long coherence times and high-fidelity gate operations and measurements~\cite{van2022quantum}, so protocols can run in a distributed fashion.
However, such requirements make these layered stacks very challenging to deploy on existing quantum networking hardware.
Thus, Pirker et al.~\cite{Pirker2025} proposed a radical shift from layered architectures to a resource-centric, task-based model, in which users define high-level objectives that are fulfilled through distributed workflows---called ``sagas''---that operate directly on quantum network resources (e.g., quantum channels, classical messaging, and shared entanglement). 
Cacciapuoti and Caleffi followed up in the same direction with a quantum-native organizational principle based on dynamic composition, which replaces static layering with a distributed orchestration fabric driven by in-band control and the local state of quantum network nodes~\cite{cacciapuoti2026}.

While it is important to define architectures for a quantum internet, near-term quantum networks will likely rely on centralized architectures~\cite{Chung2022,wehnerscheduling,Schon2024,quantnet-qce25}.
Motivated by the flexibility of the resource-centric, task-based architecture~\cite{Pirker2025}, in this paper we implement a centralized version of this architecture's control plane in SeQUeNCe~\cite{sequence}, a discrete-event simulator of quantum networks.
We evaluate our central controller using a diverse set of physical network topologies (i.e., star, bottleneck, grid, and caveman) and traffic patterns.
Our results reveal that the caveman and grid topologies outperform the star and bottleneck topologies at delivering small request delays for the same traffic patterns due to the availability of alternative paths for serving more user requests in parallel. However, these topologies also deliver a significant fraction of requests with higher delays than any found in the star and bottleneck topologies.
Furthermore, our simulation results lead to the conclusion that our centralized implementation of the resource-centric, task-based architecture is robust and performant in situations of high load in quantum networks. Our results showcase the viability of the resource-centric task-based quantum network control framework~\cite{Pirker2025} for various kinds of quantum network topologies and traffic load patterns, and underline its suitability for high load situations. Our SeQUeNCe implementation of the centralized, resource-centric, task-based controller is open-sourced on GitHub~\cite{repo}.

The rest of the paper is organized as follows.
Section~\ref{sec:background} provides the necessary background and motivation for this work, while Sections~\ref{sec:arch} and~\ref{sec:system_model} describe the resource-centric, task-based architecture in detail and present our system model, respectively.
In Section~\ref{sec:methods} we describe our implementation in SeQUeNCe and the methodology to execute our simulations.
Section~\ref{sec:results} discusses our results.
Finally, Section~\ref{sec:conclusion} concludes the paper.

%% file: sections/background.tex
\section{Background} \label{sec:background}
In classical networks, devices operate by adopting a packet-switching paradigm that eventually resulted in the famous Open Systems Interconnection (OSI) model~\cite{Zimmermann88}. Within this model, devices organize their responsibilities hierarchically into seven layers, each of them having its own protocol (suites). Each protocol wraps additional information around the data sent through the network, and devices operating on layers unwrap packets until their corresponding layer for processing. 

The main goal of a quantum network is different from classical packet-switching networks. Unlike classical networks, devices in a quantum network can share entanglement, a vital resource that enables many distributed quantum applications. When generating entanglement for applications, quantum networks follow one of two approaches: either building up entanglement \cite{Campbell2007,RepeaterNet1,RepeaterNet4} on demand (bottom-up approach) or consuming a pre-shared, general-purpose, entangled quantum state, referred to as network state, which is generated before any request is issued~\cite{Pirker_2018,Freund2025} (top-down approach). Both approaches have advantages and disadvantages. For instance, in a bottom-up approach no overly expensive, general-purpose, entangled state must be established and kept alive, whereas the time-to-serve is significantly longer compared to a top-down approach. From these approaches, three kinds of quantum network stacks have emerged in the last decade \cite{VanMeterStack1,WehnerStack1,WehnerStack2,WehnerStack3,Pirker_2019}, all with the goal of generating entanglement (bi- or multi-partite) between network nodes.

The stack model proposed by Van Meter et al.~\cite{VanMeterStack1} establishes at its lowest layers Bell-states with the direct neighbors of a quantum network node. On top, it nests layers that apply entanglement purification and entanglement swapping, ultimately serving the application layer which consumes the generated Bell-states.  Similarly, the model of \cite{WehnerStack1,WehnerStack2,WehnerStack3} also generates Bell-states between neighboring nodes at its lowest layer, here by utilizing the Midpoint-Heralding-Protocol (MHP). The layers on top facilitate entanglement attempts and long-distance Bell-states. Once the long-distance Bell-state is ready, the transport layer consumes it for quantum teleportation \cite{Bennett96}. In contrast to the previous two models, Pirker and D\"ur's model~\cite{Pirker_2019} aims at distributing long-distance multipartite entanglement instead of solely Bell-states. At its lowest layer it also generates short-distance entanglement, but the second layer generates long-distance, multi-partite entanglement (network states). The next layer utilizes the network state shared between the devices of the network to fulfill requests, such as generating graph states in the quantum network. Lastly, the network layer connects quantum networks again via multipartite entangled states. 

Entanglement, in general, corresponds to the central, desirable resource in a quantum network, acknowledged by all stack models \cite{VanMeterStack1,WehnerStack1,WehnerStack2,WehnerStack3,Pirker_2019}. When implementing a quantum network stack, one faces major challenges arising from the very nature of entanglement, making it difficult to organize a quantum network hierarchically. First, a hierarchical organization imposes a strict layering in terms of how to accomplish a goal, because each layer has a dedicated responsibility. When passing information from higher to lower layers, each layer introduces processing times and copies of classical information (as layers are independent), thus increasing waiting times. These times unavoidably degrade the quality of entangled states kept in memory (due to decoherence), directly impacting the final fidelity of the quantum state. Second, entanglement corresponds to a resource in a quantum network which naturally occurs at any given stage of request processing, spreading potentially through multiple layers. We point out that actions performed by different layers may combine (or fuse) separate entanglement resources by swapping operations or other entangling gates. Therefore, associating and tying entanglement to a particular hierarchical layer inherently limits its usability in a quantum network. Third, some entangled quantum states appear in a transient manner as part of serving a request, and may not require full tracking in a hierarchical organization due to their transient nature. Fourth, the decoherence times strictly limit the amount of tolerable waiting times in general, and hence a streamlining and optimization of task processing times is obligatory to achieve the highest possible fidelity of quantum states.

It must be noted that centralized approaches to quantum network control have been proposed also in the past. For example, scheduling entanglement generation attempts in discrete time-slots in quantum networks was considered in \cite{wehnerscheduling}. The work of \cite{Schon2024,quantnet-qce25} introduces a centralized controller orchestrating and coordinating all activities of quantum network nodes. In \cite{Chung2022} two layers were introduced for the quantum network control plane, namely a device control functions layer in the bottom (interfacing with the infrastructure plane) and a network control function layer on top (interfacing with the service plane).

%% file: sections/architecture.tex
\section{Architecture}\label{sec:arch}
\begin{figure}
    \centering
    \includegraphics[trim={0.5cm 1cm 0.5cm 1cm}, clip, width=\columnwidth]{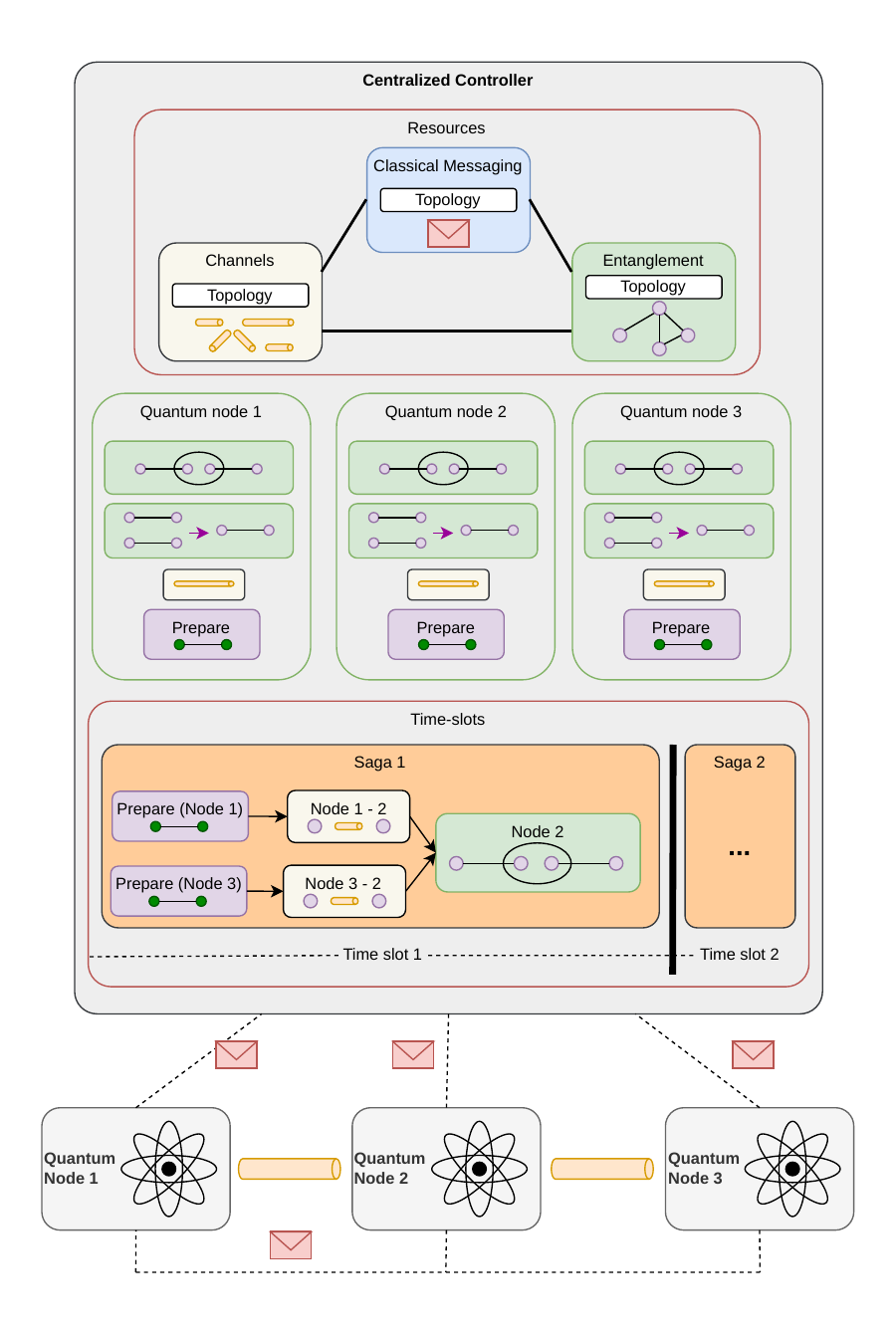}
    \caption{The resource-centric, task-based approach to quantum network control using a centralized controller. The controller keeps track of the resources in the network (channels, entanglement, classical messaging), and it is aware of the executable tasks (capabilities) of the nodes in the network (such as entanglement swapping, entanglement purification, etc.). It derives distributed workflows (sagas) to accomplish objectives, and schedules them for execution in the network.}
    \label{fig:architecture}
\end{figure}

To overcome the shortcomings of quantum network layered stacks, Pirker et al.~\cite{Pirker2025} proposed a resource-centric, task-based approach to quantum network control. In this framework, users or applications pose objectives to the control plane of the network. 
It must be noted that the architecture is agnostic of how the control plane is implemented.
In principle the control can be both distributed across the nodes of the network or centralized into a single controller. In the following, we describe the ideas of \cite{Pirker2025} for the use-case of a centralized controller, but the same principles apply in a distributed setting. Figure~\ref{fig:architecture} shows the overall architecture.

An objective abstractly characterizes what a user demands from the network, like for example generating a Bell-state of certain fidelity between two remote network nodes. In fact, an objective contains only the description of what should be achieved by the network, lacking any actionable instructions about how to accomplish it. The entire framework, at its heart, builds around the resources in a quantum network. The work of \cite{Pirker2025} identified three kinds of resources, namely (quantum) channels, entangled quantum states and classical messaging between the nodes. In addition to the resources, the centralized controller is also aware of the capabilities of each quantum network node. These capabilities correspond to elementary tasks executable by a quantum network node, such as applying quantum operations, performing entanglement swaps, sending qubits through a channel, or running an entanglement distribution or purification protocol. For a complete list, we refer the reader to \cite{Pirker2025}. The responsibility of the centralized controller is to derive a distributed workflow for achieving the objective, referred to as saga. When deriving a saga, the controller optimizes the composition of a saga into tasks by taking into account the available resources of the quantum network. After deriving a saga, the centralized controller schedules the saga for execution in the network by its nodes. 

\begin{figure}
    \centering
    \includegraphics[trim={0.5cm 1cm 0.5cm 1cm}, clip, width=\columnwidth]{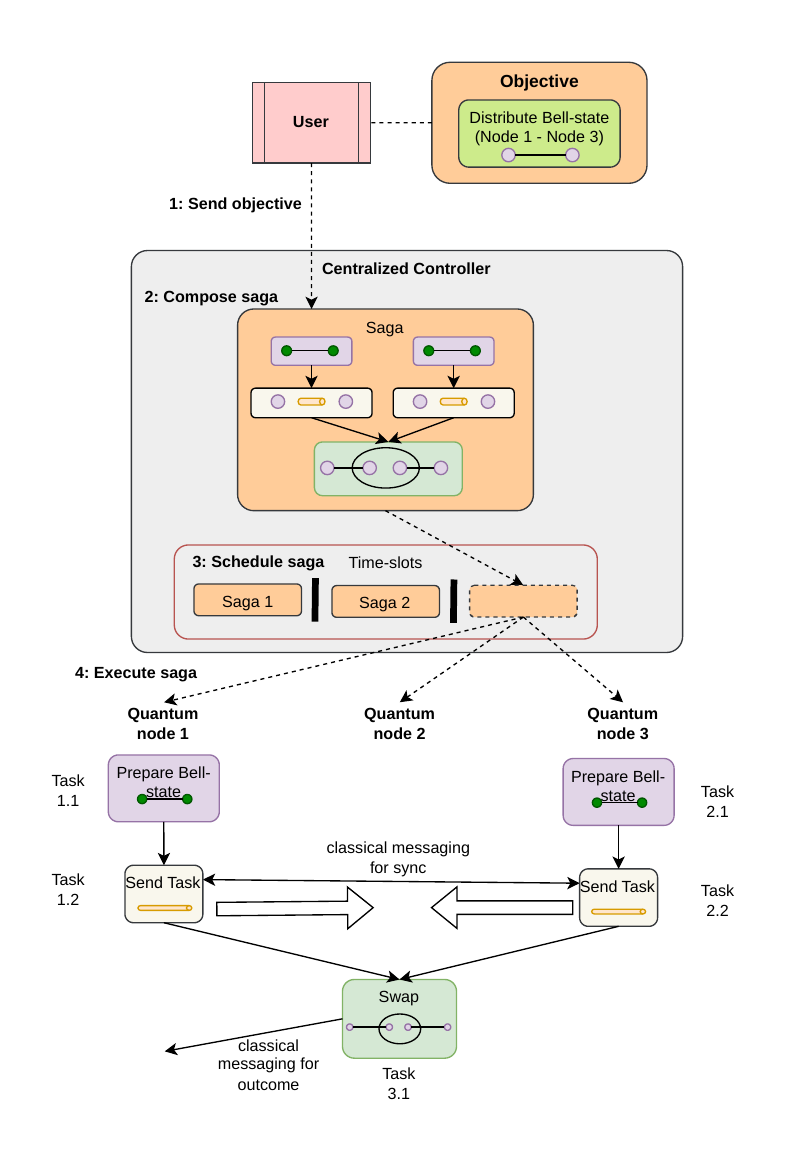}
    \caption{The control flow for quantum networks adopting a centralized, resource-centric, task-based approach to quantum network control. (1) A user sends the objective to the centralized controller which in turn (2) derives a saga composed of elementary tasks, exploiting the available resources of the quantum network. Once ready, (3) the controller schedules the saga for execution in the network in a time-slot. When the network advances to that time-slot, (4) the centralized controller initiates the saga execution by the nodes of the network which ultimately achieves the objective by executing the tasks as specified in the saga.}
    \label{fig:flow}
\end{figure}

From a user's perspective, the control flow (see Fig.~\ref{fig:flow}) works as follows. \circled{1} A user of the network sends the objective to the centralized controller. \circled{2} The centralized controller consults the currently available resources in the network and derives---based on these resources---a saga in terms of executable tasks. We note that in some situations it is beneficial to consume and utilize pre-shared entanglement resources rather than create entangled states from scratch. The output of this step, namely the saga, is a concrete way of how to achieve the objective, in terms of elementary tasks utilizing the resources of the network. Once the saga is derived, \circled{3} the centralized controller schedules the saga for execution in the network. When the saga is scheduled for execution, \circled{4} the centralized controller kicks-off the saga execution by distributing the saga in terms of its tasks to all involved nodes of the network. We note that alternatively the centralized controller could also distribute the saga (and its tasks) immediately after derivation and prior to its execution to the nodes. In this situation the controller sends only the start signal for the saga when the slot of its execution is due to the nodes. Consequently, the network nodes execute the tasks according to the specified saga, thereby achieving the objective.

\begin{figure}
    \centering
    \includegraphics[trim={0.5cm 0.25cm 0.5cm 0.25cm}, clip, width=\columnwidth]{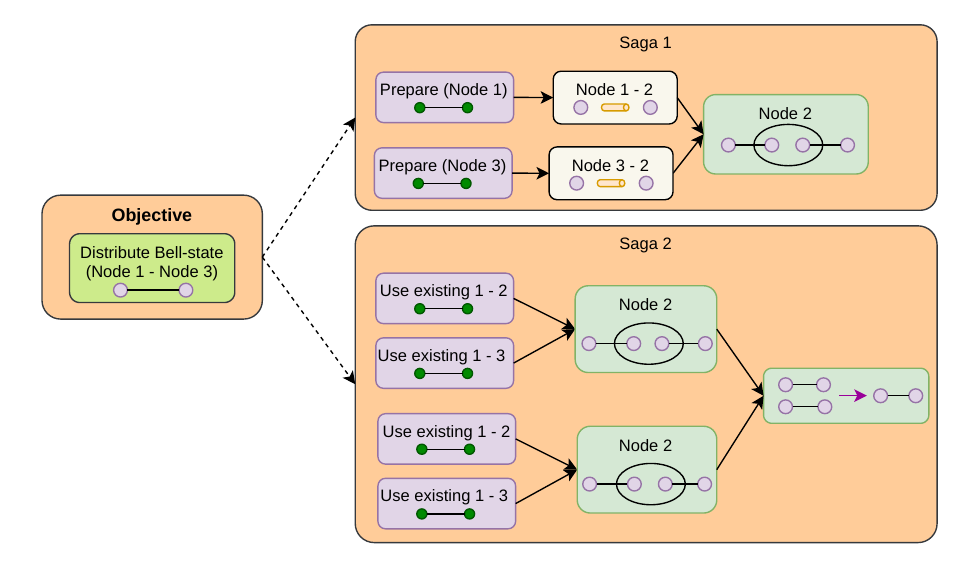}
    \caption{To achieve an objective, in principle several sagas could be used. For example, when distributing a Bell-state between Node 1 and Node 3, Saga 1 could be used (prepare local Bell-states, send a half to Node 2 which performs an entanglement swap). Alternatively, pre-shared entanglement between Node 1 and Node 2 and Node 1 and Node 3 could be swapped by Node 2, followed by performing an entanglement purification step. Both sagas achieve the same objective, namely a Bell-state between Node 1 and Node 3.}
    \label{fig:alternatives}
\end{figure}

We point out, as also described in \cite{Pirker2025}, that the composition of a saga to achieve an objective depends on the availability of resources in the network, which may vary over time. For example, in one situation it is preferable to consume pre-shared entangled states in the network over generating new ones from scratch by using the quantum channels in the network. 
Figure~\ref{fig:alternatives} illustrates this example of how one objective can be fulfilled by two different sagas. This observation lies at the heart of the resource-centric, task-based approach.

%% file: sections/system_model.tex
\section{System Model}\label{sec:system_model}
We consider an arbitrary quantum network topology described by an undirected graph, $G(V,E)$, where $i\in V$ represent quantum routers and $(i,j)\in E$ represent optical channels. The quantum network has a central controller and $N = |V|$ quantum routers, each with $m$ quantum memories. All nodes, including the controller, communicate over an all-to-all reliable classical network. Users generate objectives and send them to the controller. 

We assume that all quantum channels are modeled by a depolarizing channel with equal weights for $X$, $Y$, and $Z$ errors determined by parameter $p_e$.
We additionally restrict the form of entangled states $\rho$ within our work to Werner states of the form
\begin{equation}\label{eqn:werner}
    \rho = F\Phi^+ + \frac{1-F}{3}\left(\Phi^- + \Psi^+ + \Psi^-\right)
\end{equation}
where $F$ represents the state fidelity and our target entangled state is $\Phi^+$. For quantum memory storage, the time dependence of the channel can be described by
\begin{equation}\label{eqn:decay}
    p_e(t) = e^{-t/\tau}
\end{equation}
for time $t$ and decoherence time $\tau$. If a Werner state with fidelity $F_{in}$ is stored in two quantum memories with identical coherence times $\tau$ for time $t$, the final state is still a Werner state but with a degraded fidelity~\cite{zang2023entanglement}
\begin{equation}\label{eqn:fidelity}
    F(t) = F_{in}e^{-2t/\tau}+\frac{1-e^{-2t/\tau}}{4}
\end{equation}

For any two-qubit gate operation, we assume that the probability the two qubits are depolarized after the operation is $p_g$. Additionally, for a single-qubit measurement, we assume the classical result is incorrect with probability $p_m$. We then define the gate fidelity to be $1-p_g$ and the measurement fidelity to be $1-p_m$.

Using this error model and two Werner states as input with fidelities $F_1$ and $F_2$, we describe the entanglement swapping output fidelity using a noisy CNOT gate and imperfect single-qubit measurement operations. The output fidelity is described as~\cite{zang2023entanglement}:
\begin{align}
    F_\mathrm{swap,W} & = \frac{p_{\text{g}}}{4} + (1-p_{\text{g}})\left[(1-p_{\text{m}})^2\left(F_1F_2 + 3e_1e_2\right)\right.\nonumber \\
    & \left. +p_{\text{m}}(p_{\text{m}}-2)\left(F_1e_2 + e_1F_2 + 2e_1e_2\right)\right],
    \label{eq:sequence-fidelity-verification}
\end{align}
where $e_i=(1-F_i)/3$ are defined to simplify the notation.

The controller composes each objective into a saga, also referred to as \emph{reservation}, which describes the resource requirements together with a set of actionable items (planned tasks) for each objective across the quantum network. A centralized offline scheduling protocol adjusts sagas to resolve potential conflicts. Next, we describe objectives, sagas, and the scheduling protocol. 

We consider that each user objective requests $n_p$ entanglement (Bell) pairs between an arbitrary source-destination pair $i\in V$ and $j\in V$ using $k$ memories for a duration $\Delta t$.  We represent each objective using the tuple:
\begin{equation}\label{eqn:res}
    r = (i, j, t_a, t_e, k, F, n_p, \iota, p)
\end{equation}
where $(i,j)$ are the endpoints, $t_a$ is the arrival time, $t_e=t_a+\Delta t$ is the end time, $F$ is the target fidelity, $\iota$ is the identifier, and $p$ is the priority. 

The offline scheduling protocol is summarized in Algorithm~\ref{alg:offline-scheduling}, and is as follows. Objectives are pushed onto a min-heap $Q$ ordered by $(p, t_a, \iota)$. The central controller processes each objective in the min-heap iteratively, until the min-heap is empty. The central controller first computes the required path $D_{ij} \subseteq E$ using Dijkstra's shortest path algorithm, laying the foundation for the objectives' saga. Then for each objective and for each node $v$ in the path $D_{ij}$, the central controller attempts to reserve quantum memories by scheduling the saga to the nodes' timecards $T_v$ in the interval $[t_a,t_e]$, where source and destination nodes $(i,j)$ need $k$ memories, and intermediate nodes need $2k$ memories. The controller assigns available memories and sequentially checks each time card for scheduling conflicts using binary search until the requested memory number is satisfied. If a conflict is identified and a saga is rejected, i.e., fails to schedule due to insufficient available quantum memories, then the start time $t_a$ is delayed by one time-slot scaled by the saga's priority, i.e., $(p+1)\Delta t$. This adjustment ensures that all sagas will be scheduled in the simulation. Because the min-heap is ordered by priority, sagas with lower priority will be delayed longer in favor of early scheduling of high priority sagas. When each saga is scheduled to the timecards, the central controller will transmit an approval message and the corresponding timecards to each node in the path $D_{ij}$. The path itself together with further parameters can be interpreted as a set of \emph{planned tasks} for how to accomplish the objective.

\begin{algorithm}
\caption{Centralized Offline Scheduling}
\label{alg:offline-scheduling}
\begin{algorithmic}[1]
\Require Graph $G(V,E)$; set of $n_o$ objectives $R$, each with a tuple~\eqref{eqn:res}; node timecards $\{T_v\}_{v \in V}$
\Ensure Approved sagas $A$, each with assigned path $D_{ij}$ with nodes from $V$ and optical links from $E$
\State $Q \gets \Call{heapify}{R}$ ordered by $(p, t_a, \iota)$
\While{$Q \neq \emptyset$}
\State $r \gets \Call{Pop}{Q}$
\State Let $(i, j)$ be the endpoints of $r$ in~\eqref{eqn:res} 
\State $D_{ij} \gets \Call{Dijkstra}{G, i, j}$
\State $S\gets []$
\For {$v\in D_{ij}$}
\State $k^\prime \gets k$ \textbf{if} $v \in \{i, j\}$, \textbf{else} $2k$
\If {$T_v$ assigns $k^\prime$ memories to $r$}
\State $S.\mathrm{append}(v)$
\Else
\For{$u\in S$} 
\State unassign $r$ from $T_u$
\EndFor
\State $\Call{Backoff}{t_a}$
\State $\Call{Push}{Q,r}$
\State \textbf{break}
\EndIf
\EndFor
\If{$|S| = |D_{ij}|$} $A.\text{append}(r)$ \EndIf
\EndWhile
\State \Return $A$
\end{algorithmic}
\end{algorithm}

We summarize the terminology of the system model in Table~\ref{tab:terminology}.

\begin{table}[h]
    \centering
    \caption{Terminology}
    \begin{tabular}{l l}
    \toprule
        Definition & Representation \\
    \midrule
        Number of Objectives& $n_o$ \\ 
        Number of Bell pairs per Objective & $n_p$ \\
        Objective & $r$ \\
        Source & $i$ \\
        Destination & $j$ \\ 
        Arrival Time & $t_a$ \\
        End time & $t_e$ \\
        Required memories & $k$ \\
        Target fidelity & $F$ \\
        Priority & $p$ \\
        ID & $\iota$ \\
        CNOT Gate Error & $p_g$ \\
        Measurement Error & $p_m$ \\
    \bottomrule
    \end{tabular}
    \label{tab:terminology}
\end{table}

%% file: sections/methodology.tex
\section{Methodology} \label{sec:methods}
\begin{figure}[htbp]
    \centering
    \includegraphics[width=\columnwidth]{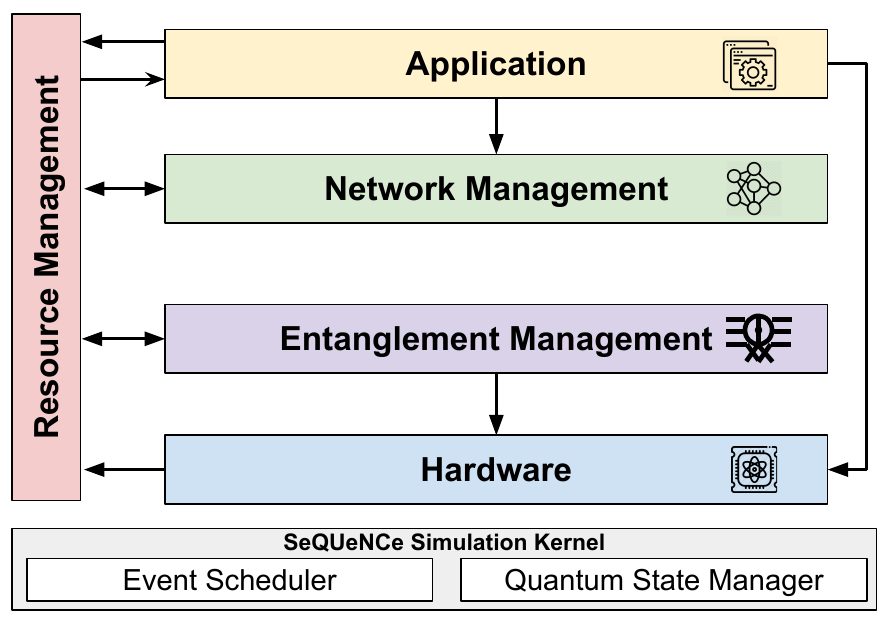}
    \caption{SeQUeNCe modular architecture of a node and the simulation kernel. Reproduced under the terms of the CC-BY license~\cite{sequence}. Copyright 2021, The Author(s).}
    \label{fig:sequence}
\end{figure}

Following the architecture described in Sec.~\ref{sec:arch} we implement the system model described in Sec.~\ref{sec:system_model} using SeQUeNCe, a modular discrete-event simulator of quantum networks.
Figure~\ref{fig:sequence} shows the modular architecture of a quantum network node in SeQUeNCe along with the simulation kernel. 
By default, SeQUeNCe implements a distributed resource reservation scheme. The main module of focus is the network manager, which contains routing, scheduling, and forwarding logic. The network manager accomplishes scheduling by maintaining a local set of memory timecards which contain time-series utilization information for each quantum memory. Given a resource reservation request, the network manager reserves resources using a distributed RSVP protocol among nodes in a path.
At its core, the RSVP protocol performs scheduling and path calculation locally given a reservation. 

This work introduces a centralized version of SeQUeNCe's network manager in which we co-locate the quantum memory timecards with the central controller. This enables the controller to perform offline scheduling and path calculation using the full timecard information as described in Sec.~\ref{sec:system_model}.
Our SeQUeNCe implementation of the centralized, resource-centric, task-based controller is open-sourced on GitHub~\cite{repo}.

Table~\ref{tab:sequence} maps the terminology used in the architecture to naming conventions in SeQUeNCe.
Specifically, the centralized, task-based architecture is implemented in SeQUeNCe as follows:
\begin{enumerate}
    \item A user generates an \emph{objective} which corresponds in this work to distributing Bell pairs between two nodes in the quantum network. The user communicates this \emph{objective} to a central controller using a classical channel.
    \item From this \emph{objective}, the central controller composes a \emph{saga} that is implemented in SeQUeNCe using a \texttt{reservation}. Thus, \emph{sagas} encode a path in the network corresponding to a list of \emph{planned tasks} for how to achieve the objective together with information from which nodes determine \emph{(actionable) tasks} that are not yet executed, but planned for execution.
    \item The central controller communicates \texttt{reservations} to the involved nodes over classical messages.
    \item Nodes execute the saga, i.e., \texttt{reservation}, by turning the \emph{planned tasks} from the saga into concrete items as \emph{(actionable) tasks} using the \texttt{action-rule-set} table from SeQUeNCe's resource manager.
\end{enumerate}
Actionable tasks in SeQUeNCe specifically include the probabilistic generation of entanglement between adjacent nodes with imperfect fidelity $F$ according to Equation~\ref{eqn:werner} and entanglement swapping with additional fidelity overhead described by Equation~\ref{eq:sequence-fidelity-verification}. Between tasks, entanglement fidelity decays according to Equation~\ref{eqn:fidelity}.

To generate user requests, we sample the $n_o$ objectives from a traffic pattern. 
Specifically, we create a traffic matrix $M$, where $M_{ij}\sim\text{Poisson}(\lambda), \forall i\neq j; i,j \in \{1,\ldots,N\}$, and $M_{ii}=0, \forall i\in\{1,\ldots,N\}$, that captures the source-destination pairs from users' objectives. From $M$, we generate the probability mass function $P_{ij}=M_{ij}/\sum_{i'\in V,j'\in V,i'\neq j'}M_{i'j'}$. Then, objectives are generated sequentially by sampling a source-destination pair from $P_{ij}$, its priority from $p\sim \text{U}\{0, 1, 2\}$, and its inter-arrival time (i.e., the time period between two consecutive objective arrivals to the central controller) from exp$(1/\lambda)$.

\begin{table}[htbp]
    \caption{Mapping between architecture and SeQUeNCe implementation terminology}
    \centering
    \begin{tabular}{l l}
        \toprule
        Architecture~\cite{Pirker2025} & SeQUeNCe \\
        \midrule
        Objective & User Request (see Eq.~\eqref{eqn:res}) \\
        Saga & Reservation and path \\
        Task & Rules \\
        Quantum Channel & Quantum Channel \\
        Classical Channel & Classical Channel \\
        Memory & Memory \\
        \bottomrule
    \end{tabular}
    \label{tab:sequence}
\end{table}

We evaluate our central controller using the star, bottleneck, grid, and caveman topologies (see Fig.~\ref{fig:topos}). The following subsections detail our experimental methodology including our choice of topologies and simulation parameters. 

\subsubsection{Topologies}
\begin{figure}[t]
\centering
\subcaptionbox{Grid}{\includegraphics[width=0.49\columnwidth]{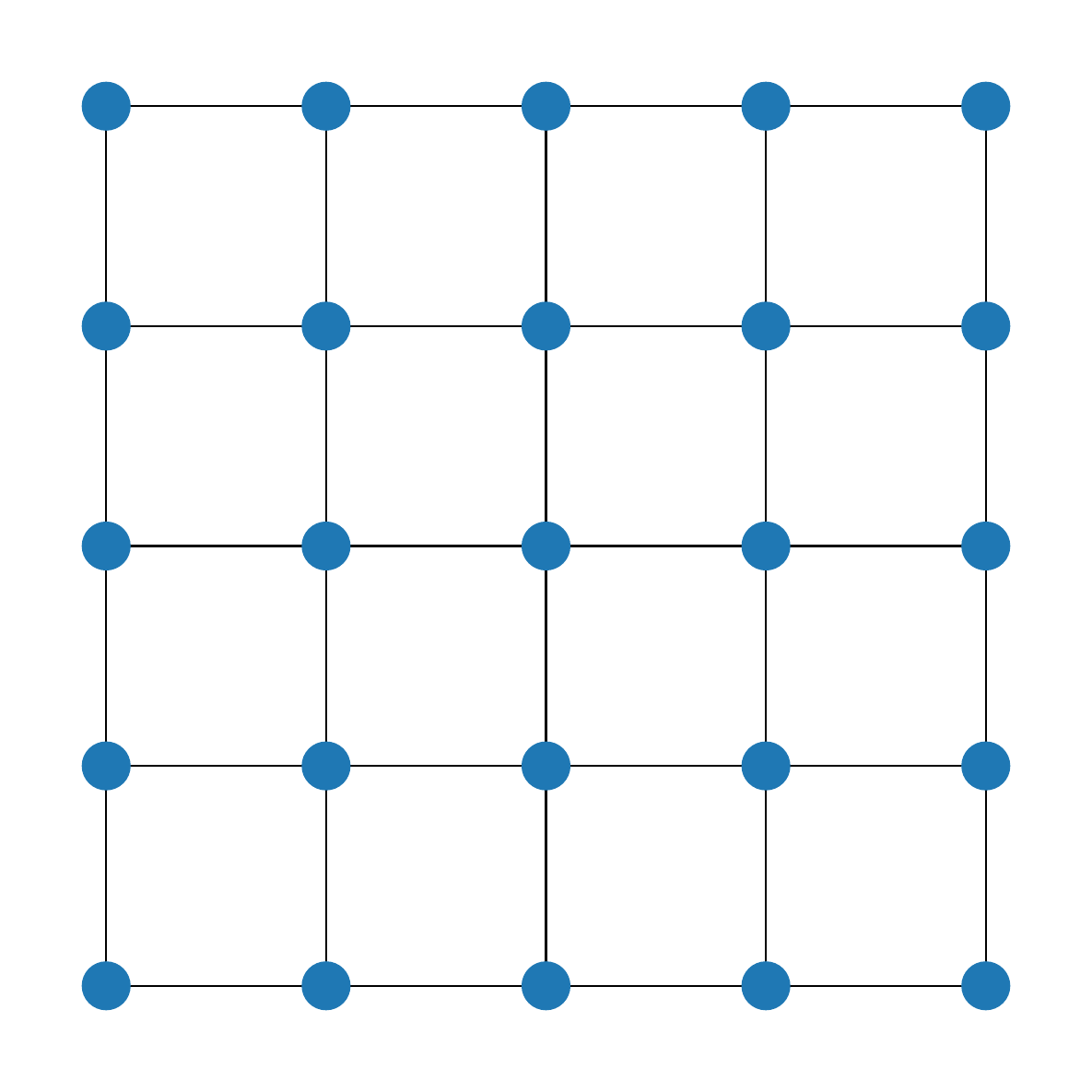}}%
\subcaptionbox{Caveman}{\includegraphics[width=0.49\columnwidth]{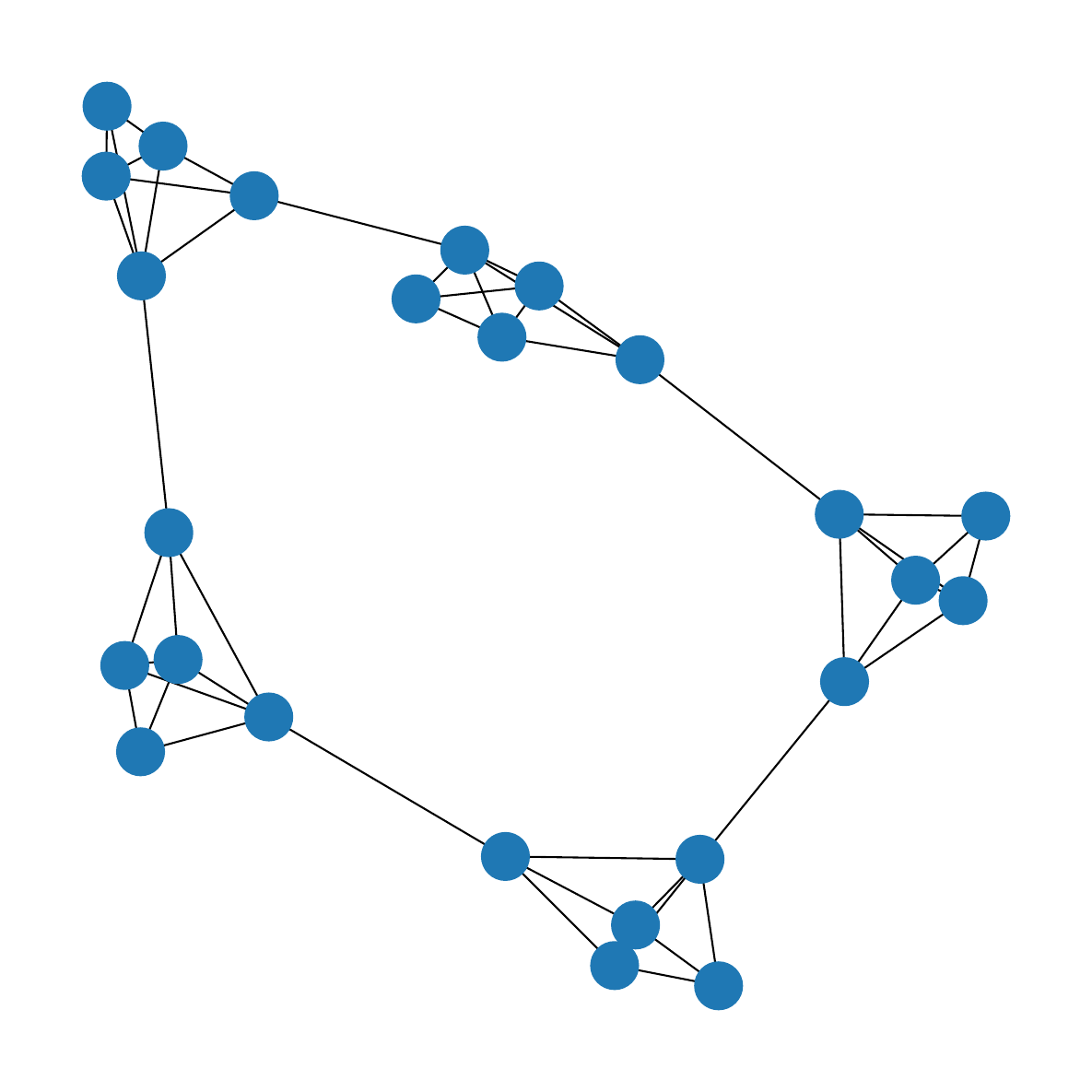}}\\
\subcaptionbox{Bottleneck}{\includegraphics[width=0.49\columnwidth]{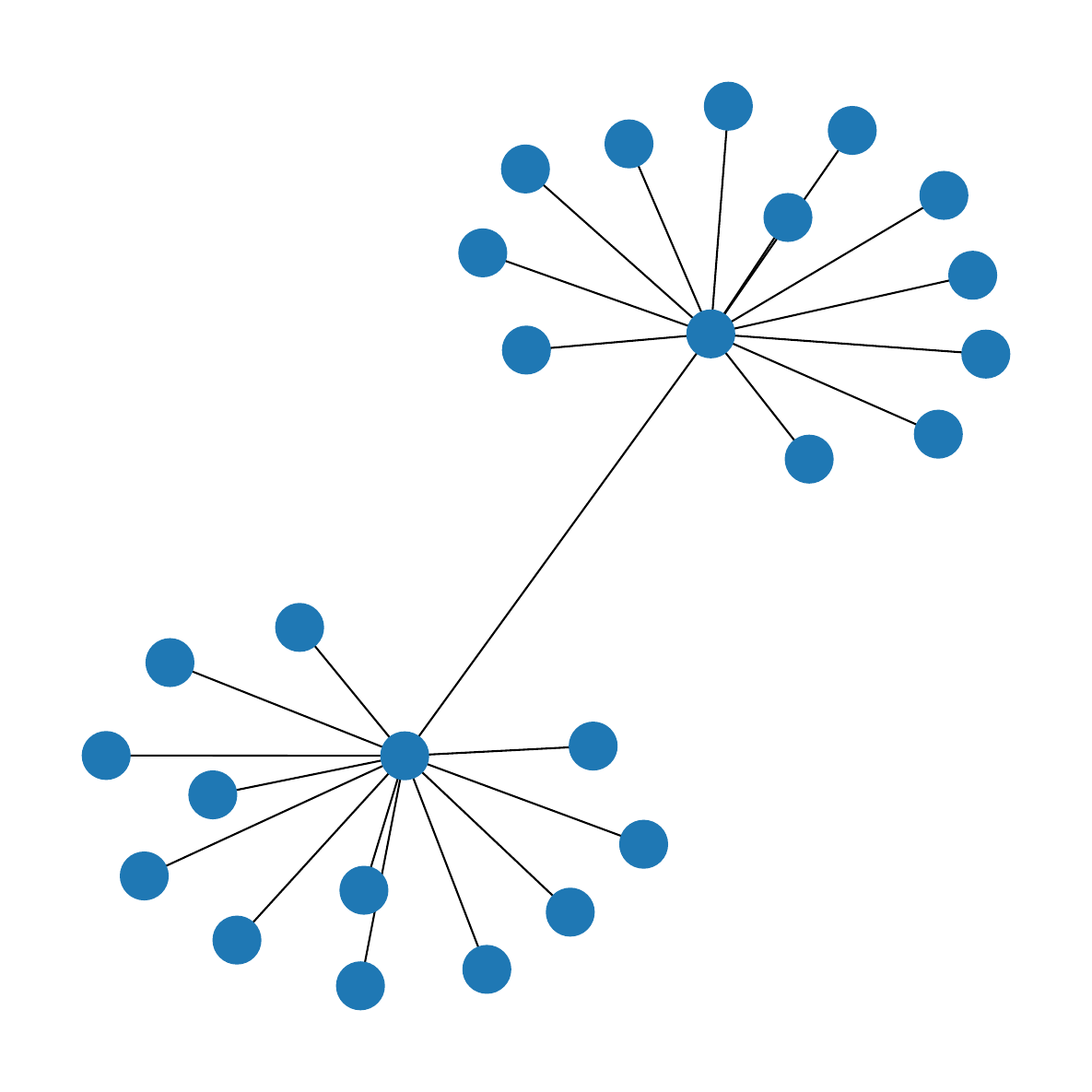}}%
\subcaptionbox{Star}{\includegraphics[width=0.49\columnwidth]{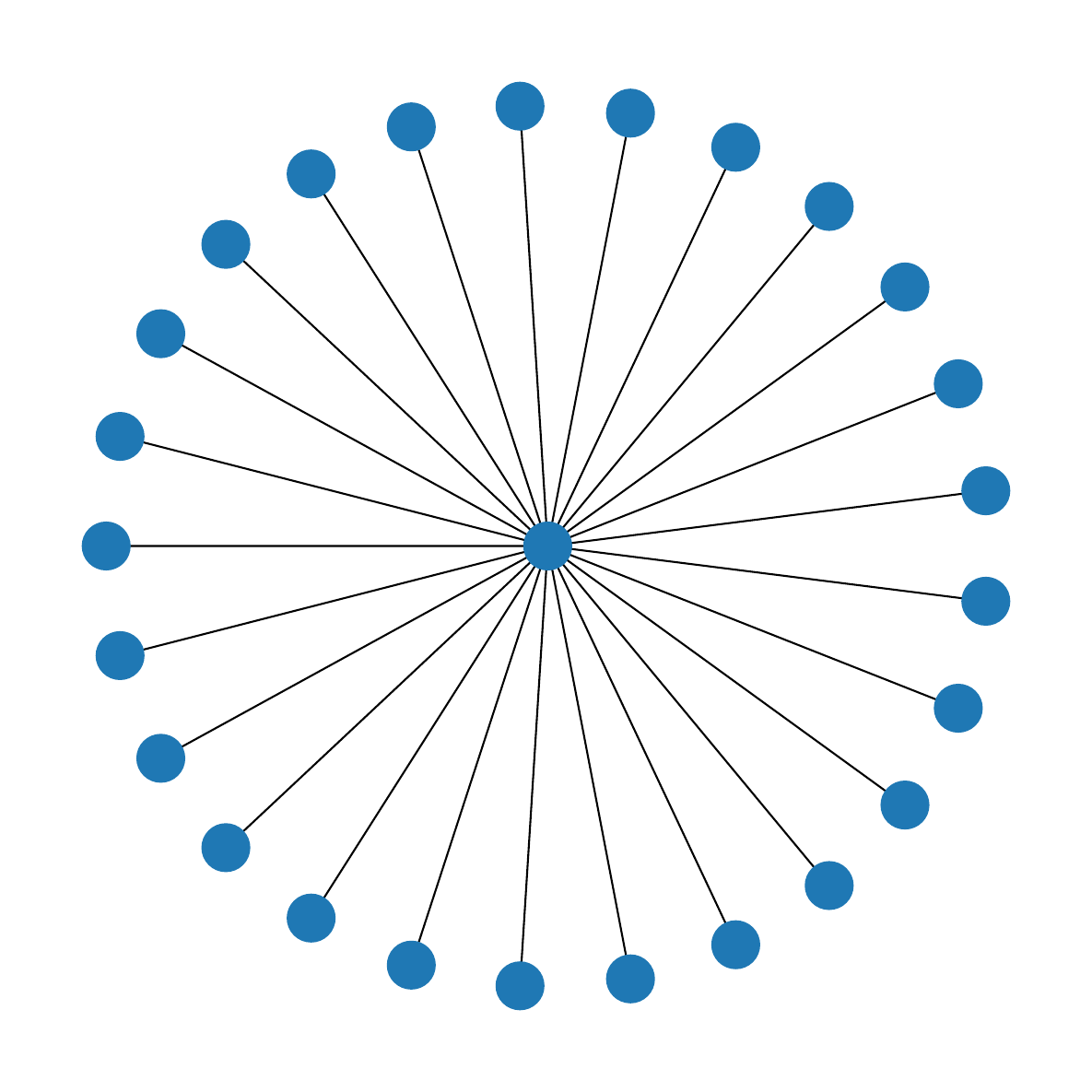}}
\caption{Network topologies.}\label{fig:topos}
\end{figure}

In each topology we consider two types of nodes: routers and processors. Router nodes do not have the capability to act as endpoints for entanglement distribution, they can only facilitate entanglement swapping. On the other hand, processing nodes are end nodes with swapping capabilities. Regardless of node capability, SeQUeNCe treats all nodes as a Quantum Router. In each topology the graph edges are considered the quantum links, and a Bell-state analyzer (BSA) is placed in the middle of each link. Additionally, each node in the topology has all-to-all classical communication with $\qty{1}{\milli\second}$ one-way latency. We implement each topology as a NetworkX~\cite{hagberg_exploring_2008} graph, and then use a translation program to generate the SeQUeNCe configuration. We simulate the star, bottleneck, caveman, and grid topologies as shown in Figure~\ref{fig:topos}.

In the star topology, one central hub is connected to $k$ leaf nodes. The central hub is a router and leafs are processing nodes. We simulate a star graph of size $k=25$, giving 26 total nodes. In the star topology, the hub is a single point of contention, as all entanglement distribution requests must swap through it. In the bottleneck topology, two star subgraphs are joined by a single edge between their hubs. We parameterize the bottleneck topology by number of leaves on the left and right star subgraphs, $k_l$ and $k_r$, respectively. We configure a bottleneck topology with $k_l=12$ and $k_r=13$, giving 27 total nodes. All leaf nodes are processing nodes and the two hubs are router nodes. In the grid topology, 25 processor nodes are arranged in a 5$\times$5 two-dimensional grid graph (with no wrap-around).  In the (connected) caveman topology, we modify $l=5$ isolated cliques of size $k=5$ by removing one edge from each clique and using it to connect to another clique aiming to connect all of them, resulting in 25 processor nodes. The connected caveman topology approximates the QFly topology~\cite{sakuma_q-fly_2025}. 

\subsubsection{Simulation Parameters}
Table~\ref{tab:sim_params} presents our simulation parameters. For each node in a given topology, we set the number of quantum memories equal to the degree of the node. The quantum nodes in the simulation will use coherence and error probability values following the modelling in~\cite{miller_simulation_2026}. Additionally, we use default SeQUeNCe values for attenuation, speed of light, and repetition rate. The reservation overlap is evaluated for both Poisson and Bernoulli interarrival distributions. After offline scheduling is complete, each node is informed of the saga with a $\qty{50}{\milli\second}$ buffer prior to the expected start time.

\begin{table}[htbp]
    \centering
    \caption{Simulation Parameters}
    \begin{tabular}{l c}
    \toprule
    Parameter & Value \\
    \midrule
    Link length, $L$ & $10$ km \\
    Fiber Attenuation & $0.2$ dB/km \\
    Classical link One-way Delay & $\qty{1}{\milli\second}$ \\
    CNOT error probability, $p_g$ & 0.99 \\
    Measurement error probability, $p_m$ & 0.995 \\
    Memory Coherence Time, $\tau$ & $\qty{2}{\second}$ \\
    Initial Entanglement Fidelity, $F_i$ & 0.9 \\
    Repetition Rate & 10 GHz \\
    Speed of Light in Fiber, $c_f$ & $2\times 10^5\si{\kilo\meter\per\second}$ \\
    \midrule
    Number of Reservations, $n_o$ & 100, 1000, 10000 \\
    Bell pairs Requested, $n_p$ & 100 \\
    Memories per Bell pair, $k$ & 1 memory on each node\\
    Reservation Duration, $\Delta t$ & $\qty{1}{\second}$ \\
    Priority, $p$ & $\mathrm{U}\{0,1,2\}$ \\
    Path, $D_{ij}$ & Dijkstra \\
    \bottomrule
    \end{tabular}
    \label{tab:sim_params}
\end{table}

%% file: sections/results.tex
\section{Results} \label{sec:results}
After all sagas are scheduled, entanglement distribution is simulated for each request using the methodology described in Section~\ref{sec:methods}. For each topology we collect the number of delivered pairs, completion time, path length (as number of hops), time to serve a request, fidelity, average pair generation time, number of retries, conflicting nodes during scheduling, and time delay. We define the time delay as the difference between the intended start time and the scheduled time.

\begin{figure}
    \centering
    \includegraphics[width=\columnwidth]{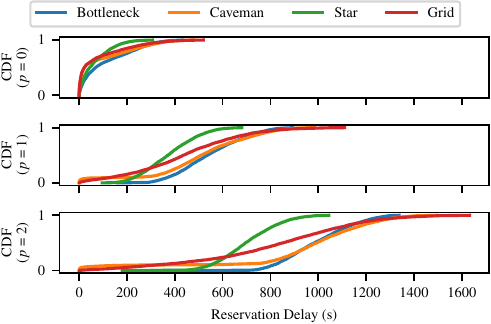}
    \caption{Cumulative distribution function of request delay for each studied topology with a queue size of 10000 requests. The graph is separated by request priority, with lowest $p$ (highest priority) requests plotted on the top sub-figure and increasing $p$ (lower priority) requests on lower subfigures.}
    \label{fig:cdf_10000}
\end{figure}

\begin{figure}
    \centering
    \includegraphics[width=\columnwidth]{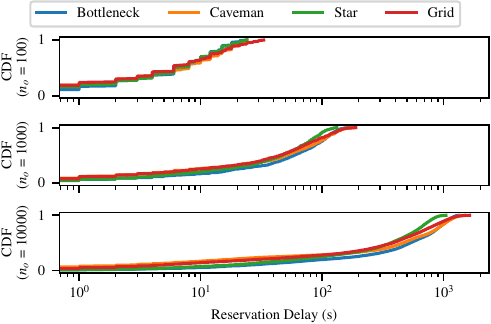}
    \caption{Cumulative distribution function of request delay for each studied topology, separated by the size of the request queue. The subgraphs show the cumulative distribution function of request delay for queue sizes $n_o = \{100, 1000, 10000\}$.}
    \label{fig:queue_change}
\end{figure}

\begin{figure}
    \centering
    \includegraphics[width=\columnwidth]{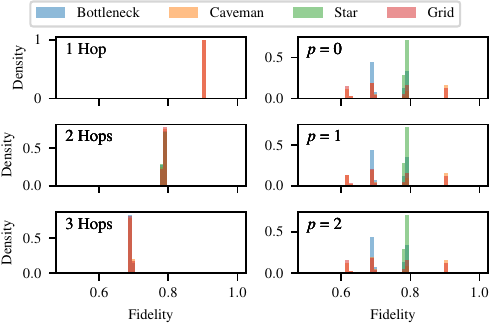}
    \caption{Histogram of the simulated entanglement fidelity over all requests. The distribution is separated by request path length (as number of hops) for the left subfigures, and separated by request priority for the right subfigures.}
    \label{fig:fidelity_compare}
\end{figure}

First, we simulate queues of reservation requests of size $n_o = \{100, 1000, 10000\}$, with arrival rate $\lambda = 50$. For the case of queue size 10000 (see Figure~\ref{fig:cdf_10000}), we plot the cumulative distribution function (CDF) of request delay imposed by the central controller for each topology, separated by the priority of the request. While the distribution remains similar across topologies for requests with priority 0, increasing priority reveals unique performance for each studied topology. As the queue size changes, we compare the distribution for each topology in Figure~\ref{fig:queue_change}, where the median delay value scales linearly with the queue size. Finally, we simulate each priority-sorted queue to measure the timing and fidelity of the resulting entanglement distribution. The fidelity, plotted in Figure~\ref{fig:fidelity_compare}, shows a strong correlation with the number of hops in the path between processors independent of the underlying network topology and consistent with the entanglement generation and swapping models present in SeQUeNCe. The fidelity shows no correlation with the request priority.
This demonstrates that the central resource manager implementation adds no additional performance impact to entangled pair generation within the network, aside from delaying the start time of requests with resource contention.

\begin{figure}
    \centering
    \includegraphics[width=\columnwidth]{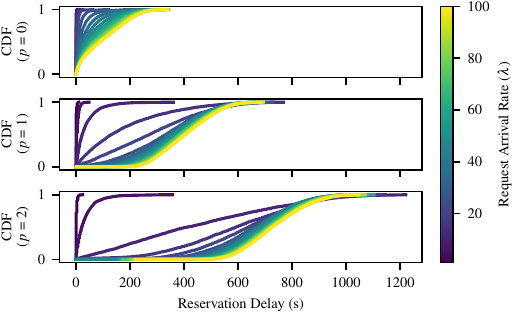}
    \caption{Cumulative distribution function of request delay with increasing arrival rate for the star network topology. The graph is again separated by request priority, with the value of $p$ for requests increasing (priority decreasing) from top to bottom.}
    \label{fig:sweep_delay_star}
\end{figure}

We also simulate 10000 reservation requests with exponentially distributed interarrival times at rate $\lambda$ while sweeping $\lambda$ across 20 linearly spaced values in the range $[1,100]$. The request delay for each arrival rate is shown in Figure~\ref{fig:sweep_delay_star} for the star network topology, separated by request priority. While the priority 0 curve interpolates smoothly between low and high arrival rates, the lower priority curves show a fast saturation to the cumulative distribution function in Figure~\ref{fig:cdf_10000} after increasing $\lambda$ beyond 20.

\begin{figure}[t]
\centering
\subcaptionbox{Grid}{\includegraphics[width=0.49\columnwidth]{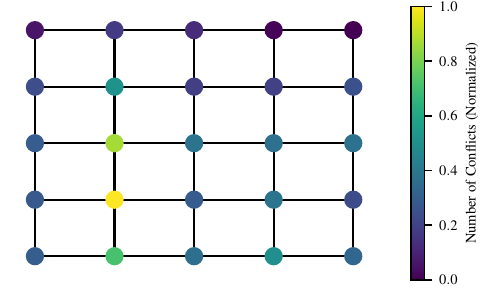}}%
\subcaptionbox{Caveman}{\includegraphics[width=0.49\columnwidth]{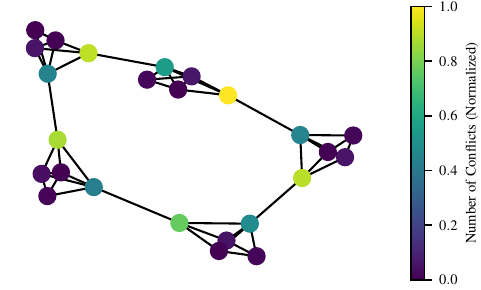}}\\
\subcaptionbox{Bottleneck}{\includegraphics[width=0.49\columnwidth]{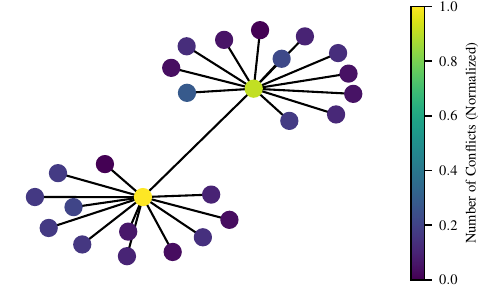}}%
\subcaptionbox{Star}{\includegraphics[width=0.49\columnwidth]{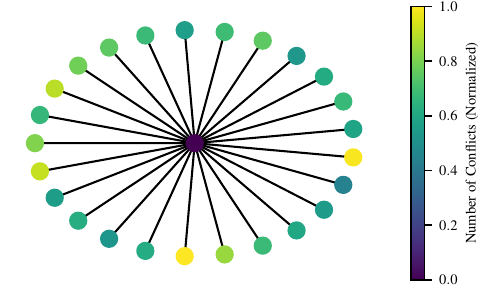}}
\caption{Normalized congestion for each node of the topologies. For each node we show the proportion of the 10000 reservations in which the node caused a scheduling conflict.}\label{fig:congestion}
\end{figure}

Figure~\ref{fig:congestion} shows the normalized proportion of the 10000 reservations in which a node caused a scheduling conflict. 
As reservations are scheduled we tracked the nodes that caused a conflict, leading to a scheduling failure in a desired time-slot.
We show that the bottleneck topology has a choke point at each of the hub nodes, and the caveman topology has high congestion at each clique connecting node. 
Interestingly, we observe that the hub node in the star topology, exhibits a low number of conflicts. 
This is the result of our methodology that sets the number of quantum memories equal to the degree of the node, a design choice grounded in lessons learned from buffer allocation in classical routers.

These results lead to the following insights. The curves of Figure~\ref{fig:cdf_10000} suggest that the caveman and grid topology both outperform the bottleneck and star topology in delivering a large fraction of short-delay requests for the centrally controlled architecture. This is due to the lack of path diversity among the bottleneck and star topologies. Each reservation in star must perform a swapping operation at the hub node. Similarly, for bottleneck, reservations must always utilize one or both hub nodes; creating congestion points. This benefit is also saturated, however, with a significant fraction of requests receiving a higher delay than any request for the star topology. Interestingly, for all three priorities the situation changes after a certain threshold, where consequently the star topology outperforms all other topologies. More surprising are the results of Figure~\ref{fig:queue_change}. The plots suggest that our centrally controlled architecture adopting a task-based approach shifts, for increasing queue sizes, the transition from low CDF values to higher CDF values along the reservation delay axis with increasing queue size. The plots suggest a linear shift. 

The results shown in Figure~\ref{fig:sweep_delay_star} yield interesting insights into the scaling of the star topology in a centralized task-based approach to quantum network control. The CDF for priority 0 shows an equidistant increase for each CDF with increasing request arrival rates. In contrast, the CDFs for priority 1 and 2 demonstrate an entirely different behavior. Both of them converge very fast to the CDF for the highest studied request arrival rate, which corresponds to a saturation. This effect is rather interesting, and together with the other results showcases that the adopted architecture performs very well under high congestion in the network for various network topologies. This underlines its suitability for different network topologies and traffic patterns.

%% file: sections/conclusion.tex
\section{Conclusion}\label{sec:conclusion}
In this work we have implemented a simulation for a centralized control plane for the resource-centric, task-based approach to quantum network control. To simulate such an architecture we have extended SeQUeNCe, a well-established quantum network simulator, accordingly. Our results reveal that the caveman and grid topologies outperform the star topology for delivering low delay requests in this framework, with an additional tradeoff of more highly delayed requests. The results of the simulation also lead to the conclusion that simulated quantum network control architecture is robust and performant in situations of high load in quantum networks. Furthermore, from the results we infer also the scalability of the framework with increasing queue sizes, as showcased for several kinds of topologies of a quantum network. In line with open-source principles, all code used in this work and data will be open-source upon publication.

For future work, we would like to explore several directions. 
First, a clear next step is to compare our centralized implementation to a distributed resource-centric, task-based network control implementation. Further, an implementation of dynamic and multipath routing would allow topologies with high path diversity such as BCube, Fat Tree, and Clos, to be evaluated, which may be more suitable for quantum data center settings.
Finally, we would like to study more complex objectives such as the composition of multipartite states that enable measurement-based quantum computation.